\documentclass[11pt,notitlepage,superscriptaddress]{revtex4-1}

\usepackage[USenglish]{babel}
\usepackage{enumerate}
\usepackage{amsmath}
\usepackage{amssymb}
\usepackage{graphicx}
\usepackage{color}
\usepackage{hhline}
\usepackage[hyperref,amsthm,amsmath,thmmarks]{ntheorem}
\usepackage{fullpage}
\usepackage{hyperref}

\theorembodyfont{\rmfamily}
\newtheorem{theorem}{\textit{Theorem} }

\newtheorem{definition}[theorem]{\textit{Definition} }
\newtheorem{corollary}[theorem]{\textit{Corollary}}
\newtheorem{lemma}[theorem]{\textit{Lemma}}
\theoremstyle{definition}
\newtheorem*{Proof}[theorem]{\textit{Proof}}

\newcommand{\lv}{\left \vert}
\newcommand{\rv}{\right \vert}
\newcommand{\la}{\left \langle}
\newcommand{\ra}{\right \rangle}
\newcommand{\ket}[1]{\lv #1 \ra}

\newcommand{\bra}[1]{\la #1 \rv}

\newcommand{\braket}[2]{\langle #1 \vert #2 \rangle}
\newcommand{\ketbra}[2]{\lv #1 \rangle \langle #2 \rv}

\newcommand{\tr}{\rm Tr}
\newcommand{\mc}[1]{\mathcal{#1}}

\begin{document}

\title{Unitary $2$-designs from random $X$- and $Z$-diagonal unitaries}

\author{Yoshifumi Nakata}
\email{nakata@qi.t.u-tokyo.ac.jp}
\affiliation{Institute for Theoretical Physics, Leibniz University Hannover, Appelstrasse 2, 30167 Hannover, Germany.}
\affiliation{Photon Science Center, Graduate School of Engineering, The University of Tokyo, Bunkyo-ku, Tokyo 113-8656, Japan.}
\affiliation{Departament de F\'{\i}sica: Grup d'Informaci\'{o} Qu\`{a}ntica, Universitat Aut\`{o}noma de Barcelona, ES-08193 Bellaterra (Barcelona), Spain.}

\author{Christoph Hirche}
\email{christoph.hirche@uab.cat}
\affiliation{Institute for Theoretical Physics, Leibniz University Hannover, Appelstrasse 2, 30167 Hannover, Germany.}
\affiliation{Departament de F\'{\i}sica: Grup d'Informaci\'{o} Qu\`{a}ntica, Universitat Aut\`{o}noma de Barcelona, ES-08193 Bellaterra (Barcelona), Spain.}

\author{Ciara Morgan}
\email{ciara.morgan@ucd.ie}
\affiliation{Institute for Theoretical Physics, Leibniz University Hannover, Appelstrasse 2, 30167 Hannover, Germany.}
\affiliation{School of Mathematics and Statistics, University College Dublin, Belfield, Dublin 4, Ireland.}

\author{Andreas Winter}
\email{andreas.winter@uab.cat}
\affiliation{Departament de F\'{\i}sica: Grup d'Informaci\'{o} Qu\`{a}ntica, Universitat Aut\`{o}noma de Barcelona, ES-08193 Bellaterra (Barcelona), Spain.}
\affiliation{ICREA--Instituci\'{o} Catalana de Recerca i Estudis Avan\c{c}ats, Pg. Lluis Companys, 23, ES-08010 Barcelona, Spain.}

%\date{resubmission April 25, 2017.}

\begin{abstract}
Unitary $2$-designs are random unitaries simulating up to the second order statistical moments of the uniformly distributed random unitaries, often referred to as Haar random unitaries. They are used in a wide variety of theoretical and practical quantum information protocols, and also have been used to model the dynamics in complex quantum many-body systems. 
Here, we show that unitary $2$-designs can be approximately implemented by alternately repeating random unitaries diagonal in the Pauli-$Z$ basis and that in the Pauli-$X$ basis. We also provide a converse about the number of repetitions needed to achieve unitary $2$-designs. These results imply that the process after $\ell$ repetitions achieves a $\Theta(d^{-\ell})$-approximate unitary $2$-design.
Based on the construction, we further provide quantum circuits that efficiently implement approximate unitary $2$-designs. Although a more efficient implementation of unitary $2$-designs is known, our quantum circuit has its own merit that it is divided into a constant number of commuting parts, which enables us to apply all commuting gates simultaneously and leads to a possible reduction of an actual execution time. 
We finally interpret the result in terms of the dynamics generated by time-dependent Hamiltonians and provide for the first time a random disordered time-dependent Hamiltonian that generates a unitary $2$-design after switching interactions only a few times.
\end{abstract}

\maketitle 

\section{Introduction}

With coherent implementations of quantum circuits becoming a reality, the question of the practical realisation of protocols in quantum information science has been a particular focus of the field in recent years. Indeed, quantum information theory itself is concerned with the evolution of quantum systems, where random processes represented by so-called \emph{Haar random unitaries} play a central role.
One of the most illustrative applications of random processes is the decoupling protocol~\cite{HaydenTutorial,DBWR2010,SDTR2013,HM14}, which provides a decoder of quantum channels and enables us to reproduce most of the known quantum capacity theorems~\cite{D2005, DW2004,HHWY2008,ADHW2009,DH2011}. Random processes are not only theoretically important but also practically useful for verifying implementations of quantum devices~\cite{EAZ2005,KLRetc2008,MGE2011,MGE2012}.
In recent years, it further turns out that random processes are a crucial key to understanding fundamental yet surprising physics in complex quantum many-body systems such as themalisation in isolated systems~\cite{PSW2006,GLTZ2006,R2008}, the information paradox of quantum black holes~\cite{HP2007,SS2008,LSHOH2013}, and quantum chaos~\cite{SS2014,RD2015,SS2015}.

While Haar random unitaries are a powerful theoretical tool in many perspectives, they cannot be efficiently implemented by quantum circuits as the number of gates required for the implementation grows exponentially in the system size. This also implies that they rarely appear in many-body systems, resulting in a lack of microscopic basis of the fundamental physics based on random process in complex quantum systems.
It is thus of crucial importance to study finite approximations of Haar random unitaries and their properties. From this point of view, a \emph{unitary design} was proposed~\cite{DLT2002,DCEL2009,GAE2007} and has been widely studied~\cite{TGJ2007,BWV2008a,WBV2008,HL2009,DJ2011,HL2009TPE,BHH2012,CLLW2015}. A unitary $t$-design is a random unitary that simulates up to the $t$th order properties of a Haar random unitary, naturally inheriting most properties of a Haar random one if $t$ is sufficiently large. 
In most applications, unitary $2$-designs are sufficient even if the designs are approximate~\cite{L2010}. 
It is also known that unitary $2$-designs can be efficiently implemented by quantum circuits such as Clifford circuits~\cite{DLT2002,AG2004,DCEL2009,CLLW2015}
%, which can be efficiently simulated even classically~\cite{AG2004}, 
and random circuits~\cite{HL2009, DJ2011}.
Some of them are already used in experiments as a standard technique to benchmark small quantum devices~\cite{RLL2009,BWCetc2011,CGJetc2013,BKMetc2014}.
Also, Hamiltonian dynamics with fully random interactions were shown to realise unitary $2$-designs~\cite{HL2009, DJ2011}, providing a possible microscopic dynamics that leads to unitary $2$-designs in complex systems. 
However, it is necessary to change interactions many times, which scales 
quadratically in the system size, before the dynamics achieves unitary 2-designs. This may take considerably long time especially in many-body systems.

This motivates the questions of whether unitary $2$-designs can be implemented by \emph{simpler} quantum circuits and also whether they can be realised by physically natural Hamiltonian dynamics in many-body systems, where the interactions vary only \emph{constant} times.
In this article, we propose a new construction of unitary $2$-designs by alternately and repeatedly applying random unitaries diagonal in the Pauli-$Z$ and -$X$ bases, and prove that it suffices to repeat them only a few times before unitary $2$-designs are achieved. We also provide the converse result, namely the necessary number of repetitions, and show that our result is tight. The converse result is obtained for the first time to the best of our knowledge, and is useful for investigating the optimality of using unitary $2$-designs in many applications~\cite{NHMW2015-2}.
We then provide a quantum circuit based on our result, where the number of gates scales quadratically in the system size. This quantum circuit is as efficient as most of the known implementations of unitary $2$-designs~\cite{DLT2002,AG2004,BWV2008a,WBV2008,TGJ2007,DCEL2009,HL2009,DJ2011}, but there exists a more efficient one~\cite{CLLW2015}, which uses a nearly linear number of gates. Our circuit has nevertheless its own merit due to its commuting property. The circuit is divided into a constant number of commuting parts, each of which is separated by the Hadamard gates, and all the gates in the commuting part can be in principle applied simultaneously. This simple structure may lead to a vast reduction in the execution time of the overall circuit. By transforming the circuit into the dynamics generated by time-dependent Hamiltonians, we construct a random many-body Hamiltonian realising unitary $2$-designs after switching the interactions a few times. 
The random Hamiltonian consists of disordered Hamiltonians with all-to-all interactions, which are similar to those in the cavity QED and thus can be implementable in actual experiments. 
Further, since disordered Hamiltonians composing the time-dependent random Hamiltonian are
considered to be a type of quantum chaos, which is expected to be a dual of quantum black holes~\cite{SS2014,RD2015,SS2015}, our result may contribute to the microscopic understandings of the quantum duality between them.

The article is organised as follows. We begin by introducing the necessary notation and definitions in Section \ref{Sect:Prelim}. The main results are presented in Section \ref{Sect:Main}. Based on the main results, we provide in Section \ref{Sect:Main} an explicit quantum circuit implementing a unitary $2$-design and a random disordered Hamiltonian realising a unitary $2$-design.
The Proofs of the main results are presented in Section \ref{Sect:Proofs}, along with statements of the necessary lemmas. The proofs of lemmas are given in Appendices.

\section{Preliminaries}\label{Sect:Prelim}

Before we state our main result, we provide a brief introduction of our notation in Subsection~\ref{SS:n} and definitions of random unitaries in Subsection~\ref{SS:rand}.

\subsection{Notation} \label{SS:n}

Throughout the paper, we use the following standard asymptotic notation. Let $f(n)$ and $g(n)$ be functions on $\mathbb{R}^+$.
We say $f(n) = O(g(n))$ if there exist $c, n_0 >0$ such that $f(n) \leq c g(n)$ for all $n \geq n_0$. 
When there exist $c, n_0 >0$ such that $f(n) \geq c g(n)$ for all $n \geq n_0$, we say $f(n) = \Omega(g(n))$.
If $f(n) = O(g(n))$ and $f(n) = \Omega(g(n))$, we denote it by $f(n) = \Theta(g(n))$.

We consider a system composed of $N$ qubits and denote by $\mc{H}_N$, the corresponding Hilbert space and by $d= 2^{N}$ the dimension of $\mc{H}_N$.
The set of bounded operators and states on a finite dimensional Hilbert space $\mc{H}$ are denoted by
$\mc{B}(\mc{H})$ and $\mc{S}(\mc{H}):=\{ \rho \in \mc{B}(\mc{H})| \rho\geq 0, \tr \rho = 1 \}$, respectively.
We also use superoperators: the most important class of superoperators in quantum mechanics is that of the \emph{completely-positive and trace-preserving (CPTP)} maps, which is also referred to as \emph{quantum channels}, because any allowed physical dynamics is represented by a CPTP map.
A CPTP map $\mc{C}$ is a linear map satisfying $\mc{C} \otimes {\rm id}_k(\rho) \geq 0$ for any $k \in \mathbb{N}$ and any $\rho \geq 0$, where ${\rm id}_k$ is the identity map acting on a $k$-dimensional Hilbert space, and $\tr \mc{C}(\rho) = \tr \rho$.

We will make use of various norms throughout the article. The $p$-norm of $X \in \mc{B}(\mc{H})$ is defined by $|\!| X |\!|_p := (\tr |X|^p)^{1/p}$ for $p \geq 1$, where $|X| := \sqrt{XX^{\dagger}}$.
The 1-norm or a \emph{trace norm} is of particular importance in quantum information processing as it provides the optimal success probability $(1+|\!| \rho - \sigma |\!|_1/2)/2$ when we would like to distinguish two quantum states $\rho$ and $\sigma$.
For a superoperator $\mc{C} : \mc{B}(\mc{H}) \rightarrow \mc{B}(\mc{H}')$, we define a family of superoperator norms $|\!| \mc{C} |\!|_{q \rightarrow p}$ ($q,p\geq 1$) and the diamond norm~\cite{KSV2002} by
\begin{equation}
|\!| \mc{C} |\!|_{q \rightarrow p} = \sup_{X \neq 0 }\frac{|\!|\mc{C}(X)|\!|_p}{|\!|X|\!|_q}, \hspace{5mm} |\!| \mc{C} |\!|_{\diamond} := \sup_k  |\!|  \mc{C} \otimes {\rm id}_k |\!|_{1 \rightarrow 1},
\end{equation}
respectively. It is known that $k \leq {\rm dim} \mc{H}$ suffices to obtain the diamond norm~\cite{KSV2002}.
Similarly to the trace norm for quantum states, the diamond norm provides the optimal success probability to distinguish two quantum channels when we are allowed to use the auxiliary systems.

\subsection{Random unitaries and their $t$-designs} \label{SS:rand}

We begin with the definition of random unitaries, before discussing their roles in quantum physics, and then explain the definition of unitary $t$-designs and their operational meanings. 

\begin{definition}[{\bf Haar random unitaries~\cite{M1990}}]
{\it Let $\mc{U}(d)$ be the unitary group
 of degree d, and denote the Haar measure (i.e. the unique unitarily
 invariant probability measure, thus often called uniform distribution) 
 on $\mc{U}(d)$ by ${\sf H}$. A \emph{Haar random unitary}
 U is a $\mc{U}(d)$-valued random variable distributed according to the Haar measure, $U \sim {\sf H}$.}
\end{definition}

\begin{definition}[{\bf Random $\boldsymbol{X}$- and $\boldsymbol{Z}$-diagonal unitaries~\cite{NM2013}}]
{\it Let $\mathcal{U}_{W}(d)$ be the set of unitaries diagonal in the Pauli-$W$ basis $\{ \ket{n}_W \}_{n=0}^{d-1}$ ($W=X,Z$), given by $\bigl\{\sum_{n=0}^{d-1} e^{i \varphi_n} \ketbra{n}{n}_W :\varphi_n \in [0, 2\pi)  \text{  for  }  n \in [0,\ldots,d-1]  \bigr\}$. 
A \emph{random $W$-diagonal unitary} $U^W$ is a $\mathcal{U}_{W}(d)$-valued random variable distributed according to a probability measure ${\sf D}_W$ induced by a uniform probability measure on its parameter space $[0,2 \pi)^d$, $U^W \sim {\sf D}_W$. }
\end{definition}

Haar random unitaries are also known as a \emph{circular unitary ensemble} in random matrix theory~\cite{M1990}, and have been used to model typical dynamics in physical systems with no symmetry.
In quantum information science, they are often used in a wide variety of protocols~\cite{HaydenTutorial,DBWR2010,SDTR2013,HM14,D2005, DW2004,HHWY2008,ADHW2009,DH2011}.
They are also turned out to be the key to understanding fundamental physics in complex quantum systems~\cite{PSW2006,GLTZ2006,R2008,HP2007,SS2008,LSHOH2013,SS2014,RD2015,SS2015}.
On the other hand, random diagonal unitaries are proposed in Ref.~\cite{NTM2012} to investigate the typical dynamics in closed systems governed by a fixed time-independent Hamiltonian. In this case, the basis of the dynamics is fixed and only phases can be randomised. 
In Ref.~\cite{NTM2012}, typical phenomena led by random $Z$-diagonal unitaries were studied especially in terms of the entangling power. It was also shown that it is relevant to thermalization phenomena in isolated classical spin systems.

Despite the usefulness of these random unitaries, they cannot be efficiently implemented by quantum circuits. This is obvious from the number of parameters to be randomised, which scales exponentially with the number of qubits.
This fact also implies that neither Haar random unitaries nor random diagonal unitaries can be realised in natural many-body systems in a realistic time scale.
Hence, it is significant to introduce approximate ones, which is called \emph{unitary designs}. A {\it unitary $t$-design} is a random unitary that approximates up to the $t$th order statistical moments of a Haar random unitary. 
To define a unitary $t$-design, let $\nu$ be a probability measure on $\mc{U}(d)$ and $\mathcal{G}_{U \sim \nu}(X)$ be a CPTP map given by $\mathcal{G}_{U \sim \nu}^{(t)}(X) := \mathbb{E}_{U \sim \nu} [ U^{\otimes t} X U^{\dagger \otimes t}]$ for any $X \in \mc{B}(\mc{H}^{\otimes t})$, where $\mathbb{E}_{U\sim \nu}$ represents an average over a random unitary $U\sim \nu$.
Then, an $\epsilon$-approximate unitary $t$-design is defined as follows.

\begin{definition}[$\boldsymbol{\epsilon}$-approximate unitary $\boldsymbol{t}$-designs~\cite{DCEL2009,HL2009}] \label{Def:Ut}
{\it Let $\nu$ be a probability measure on the unitary group $\mc{U}(d)$.
A random unitary $U \sim \nu$ is called an \emph{$\epsilon$-approximate unitary $t$-design} if $|\!|  \mathcal{G}^{(t)}_{U\sim \nu} - \mathcal{G}^{(t)}_{U \sim {\sf H}}  |\! |_{\diamond} \leq \epsilon$.}
\end{definition}

The designs are called {\it exact} when $\epsilon=0$. 
Although there are various definitions of $\epsilon$-approximate unitary $t$-designs,
most definitions are equivalent in the sense that, if $U$ is an $\epsilon$-approximate unitary $t$-design in one definition, it is also an $\epsilon'$-approximate unitary $t$-design in other definitions for $\epsilon'={\rm poly}(d^t)\epsilon$. For a more detailed explanation, we refer the reader to the Ph.D thesis~\cite{L2010}. In this paper, we use the above definition because it has the clear operational meaning that an $\epsilon$-approximate unitary $t$-design cannot be distinguished up to error $\epsilon$ from a Haar random one even if we have $t$ copies of the unitary.

\section{Main results}\label{Sect:Main}

We now present our main results. In Subsection~\ref{SS:AU2}, we provide a new construction of approximate unitary $2$-designs based on the repetitions of random diagonal unitaries.
We then provide efficient quantum circuits and random Hamiltonians for unitary $2$-designs in Subsection~\ref{SS:implemetationUn}.

\subsection{A unitary $2$-design by random $X$- and $Z$-diagonal unitaries} \label{SS:AU2}

Our strategy is to alternately and repeatedly apply random $Z$- and $X$-diagonal unitaries, which physically corresponds to alternate and repeated applications of random potentials in momentum and position spaces.
The key intuition behind our strategy comes from the fact that random $Z$-diagonal unitaries strongly randomise the system if the initial state is properly chosen~\cite{NTM2012}. This is especially observed  in terms of entanglement generated by the random unitary. In fact, a random $Z$-diagonal unitary generates extremely large entanglement, which is even higher than the typical entanglement Haar random unitaries generate~\cite{HLA2006}, if the initial state has a large support in the Pauli-$X$ basis.
It is however also true that the Pauli-$Z$ basis remains invariant under the action of random $Z$-diagonal unitaries, implying that, to fully randomise the system, it is necessary to apply another random unitary that can randomise the Pauli-$Z$ basis. This can be naturally achieved by a random $X$-diagonal unitary as it is complementary to the $Z$ one. 
Thus, it is natural to expect that alternate applications of random $Z$- and $X$-diagonal unitaries may be able to fully randomise the system. Indeed, our main result states that this intuition is correct at least up to the second order.

The process of repeating random $Z$- and $X$-diagonal unitaries $\ell$ times is described by a random unitary $U[\ell]$ given by
\begin{equation}
U[\ell] := U_{\ell+1}^Z U_\ell^X U_\ell^Z \cdots U_2^X U_2^Z U_1^X U_1^Z.
\end{equation}
where $U_i^W$ are independent $W$-diagonal unitares ($i=1,\ldots,\ell+1$, $W=X,Z$). We start and end with $Z$-diagonal ones for a technical convenience. 
Noting that applying random $W$-diagonal unitaries twice in succession is equivalent to applying only one random $W$-diagonal unitary. The $U[\ell]$ can be  equivalently expressed as 
\begin{equation}
U[\ell] =  \prod_{i=\ell}^1 U_i^{'Z} U_i^X U_i^Z. \label{Eq:un}
\end{equation}
We will use this particular expression of $U[\ell]$ in the remainder of the article. 
The $U[\ell]$ can be also represented using only random $Z$-diagonal unitaries and the Hadamard transformation $H^{\otimes N}$ on $N$ qbutis;
\begin{equation}\label{Eq:AltUn}
U[\ell] = U_{2\ell+1}^Z \prod_{i=2\ell}^{1} H^{\otimes N} U_i^Z. 
\end{equation}
From this point of view, the Hadamard gates are the only non-commuting part of $U[\ell]$. We will use this expression when we consider efficient implementations of $U[\ell]$ in Subsection~\ref{SS:implemetationUn}.

Our main result is that $U[\ell]$ approaches a unitary $2$-design exponentially quickly with increasing $\ell$. The formal statement is given by Theorem \ref{Thm:Diag2} below.
\begin{theorem}[$\boldsymbol{U{[}\ell{]}}$ is an approximate unitary $\boldsymbol{2}$-design] \label{Thm:Diag2}
A random unitary $U[\ell]$ is an $\epsilon$-approximate unitary $2$-design, where
\begin{equation}
\frac{2}{d^\ell}\left[ 1-\frac{1}{d-1}\right] \leq
\epsilon
\leq \frac{2}{d^\ell}\left[ 1+ \frac{2}{d-1}\right].
\end{equation}
\end{theorem}

Theorem~\ref{Thm:Diag2} shows that repeating random $Z$- and $X$-diagonal unitaries only a few times suffices to achieve a unitary $2$-design. 
Note that, as $\epsilon= \Theta(1/d^{\ell})$, it also provides the converse from which the number of repetitions necessary to achieve unitary $2$-designs with $U[\ell]$ can be obtained. This is in contrast to previous constructions of unitary $2$-designs~\cite{DLT2002,DCEL2009,HL2009, DJ2011,CLLW2015}, where converse statements were not obtained and whether the results are tight was not clear. The converse is needed to investigate the optimality in terms of $\epsilon$ when we use $\epsilon$-approximate unitary $2$-designs in quantum protocols. In the case of decoupling, which is one of the most important applications of unitary $2$-designs, the optimality is studied in Ref.~\cite{NHMW2015-2} and new insights have been obtained.

The significance of Theorem~\ref{Thm:Diag2} lies however in its simple implementation,
which basically comes from a fact that the random unitary $U[\ell]$ is separated into commuting (random $Z$-diagonal unitaries) and non-commuting (the Hadamard gates) parts as observed in Eq.~\eqref{Eq:AltUn}. This enables us to implement unitary $2$-designs by using random time-dependent Hamiltonians where the interactions vary only \emph{constant} times. In the following subsection, we further expand this point.

\subsection{Implementations of $U[\ell]$} \label{SS:implemetationUn}

Before we discuss about the implementations by Hamiltonian dynamics, we consider approximate implementations of $U[\ell]$ by quantum circuits. This is because a random $Z$-diagonal unitary contains an exponential number of parameters to be randomised, and its exact implementations take exponential time.
In this paper, we are interested in $2$-designs, which are the second order approximations of Haar random ones. Hence, it suffices to simulate only the second order moments of $U^Z$.
Fortunately, finite degree approximations of diagonal unitaries were studied in Ref.~\cite{NKM2014}. 
Using that results and Theorem~\ref{Thm:Diag2}, we obtain the following Theorem.

\begin{figure}[tb!]
\centering
\includegraphics[width=100mm, clip]{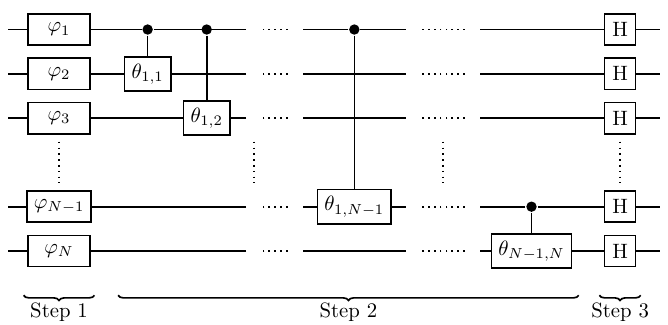}
\caption{The figure depicts a building block of the quantum circuit that implements a unitary $2$-design according to $U[\ell]$. One- and two-qubit gates in the first and the second step are given by ${\rm diag} (1, e^{i \varphi_k})$ and ${\rm diag} (1, 1,1, e^{i \theta_{l,r}})$, respectively. The phases $\varphi_k$ ($k=1, \cdots, N$) and $\theta_{l,r}$ ($l, r=1,\cdots, N$, $l \neq r$) are chosen from $\{0,2 \pi/3, 4 \pi/3 \}$ and $\{0,\pi \}$, respectively, uniformly at random. The one-qubit gates $H$ represent the Hadamard gates. All the gates in the first and the second parts are diagonal in the Pauli-$Z$ basis and can be applied simultaneously.}
\label{Fig:DUtQC}
\end{figure}

\begin{theorem}[Efficient quantum circuits for unitary $2$-designs] \label{Thm:Hma}
The quantum circuits composed of the following four steps implement $\epsilon$-approximate unitary $2$-designs (see also Fig.~\ref{Fig:DUtQC}):
\begin{enumerate}
\item Apply single-qubit phase gates ${\rm diag}_Z(1,e^{i \varphi})$, which are diagonal in the Pauli-$Z$ basis, with a random phase $\varphi \in \{0, 2\pi/3, 4\pi/3 \}$ on all qubits. 
\item Apply the controlled-phase gates $ {\rm diag}_Z(1,1,1,e^{i \theta})$ with a random phase $\theta \in \{0, \pi \}$ on all pairs of qubits.
\item Apply the Hadamard gates on all qubits.
\item Repeat the above three steps $(2 \ell + 1)$ times.
\end{enumerate}
Here, $\ell = \Theta(\frac{\log_2 1/\epsilon}{N})$,
and the total number of gates is $\Theta \bigl( N(N + \log 1/\epsilon) \bigr)$. 
\end{theorem}

In terms of the number of gates, this implementation is as efficient as most of the previously known implementations of a unitary $2$-design~\cite{DCEL2009,DLT2002,AG2004,HL2009}, but a more efficient implementation is known even for an exact unitary $2$-design~\cite{CLLW2015}.
Our implementation of a unitary $2$-design has another merit in view of commutativity of the gates, resulting in an instant property of the circuit in the sense that all the commuting parts of the circuit can be, in principle, applied simultaneously.  
This feature may lead to a practical advantage because quantum gates can be implemented by applying interactions onto two qubits and, unlike the applications of non-commuting gates, the commuting property allows us to switch on all interactions at the same time. 

This merit of the implementation is best illustrated when we interpret Theorem~\ref{Thm:Hma} in terms of the dynamics generated by time-dependent many-body Hamiltonians as given in the following corollary.

\begin{corollary}[Random Hamiltonians implementing unitary $2$-designs] \label{Cor:RHD}
Let $H_{ZX}(T)$ be a time-dependent random Hamiltonian on $N$ qubits, given by
\begin{equation}
H_{ZX}(T)= \begin{cases}
- \sum_{i=1}^N B_i^{(Z)} Z_i - \sum_{i>j} J_{ij}^{(Z)} Z_i \otimes Z_j & T \in [2 m \pi, (2m+1) \pi) \\
- \sum_{i=1}^N B_i^{(X)} X_i - \sum_{i>j} J_{ij}^{(X)} X_i \otimes X_j & T \in [(2 m+1) \pi, 2(m+1) \pi)
\end{cases}
\end{equation}
where $T$ represents time, $m=0,1,...$, and $B_i^{(W)}$ and $J_{ij}^{(W)}$ are randomly and independently chosen from $\{0, \pm 1/3\}$ and $\{0,1/2\}$, respectively.
Then, for any $T \geq T_{\epsilon}$, where $T_{\epsilon} = (5+ \lceil \frac{2 \log 1/\epsilon}{N} \rceil ) \pi + O(1/N)$ and $\lceil x \rceil$ is the ceiling function, the dynamics generated by $H_{ZX}(T)$ is an $\epsilon$-approximate unitary $2$-design.
\end{corollary}

Corollary~\ref{Cor:RHD} simply follows from the facts that the unitary generated by $H_{ZX}(T)$ at time $T=T_{\epsilon}$ is exactly the same as that in Theorem~\ref{Thm:Hma} and that applying an arbitrary random unitary onto an $\epsilon$-approximate unitary $2$-design does not change the approximation of the design.
This proof technique is rather trivial, and similar Hamiltonians can be straightforwardly obtained from any quantum circuits implementing unitary $2$-designs. 
Nevertheless, Corollary~\ref{Cor:RHD} illustrates a notable feature of $H_{XZ}(T)$, that its dynamics achieves a unitary $2$-design after switching the interactions \emph{a few} times, whereas the dynamics based on other implementations~\cite{DLT2002,AG2004,DCEL2009,HL2009, DJ2011} needs to change the interactions $O(N^2)$ times, or $O(N {\rm poly}\log N)$ times for that in Ref.~\cite{CLLW2015}, before achieving $2$-designs.
This is sorely due to the commutativity, which enables us to apply the corresponding interactions simultaneously as mentioned above. 
We also note that this small number of switches of the interactions may result in short implementations of unitary $2$-designs by Hamiltonian dynamics in physically feasible systems such as the cavity Q.E.D.

%One of the potential weak points of the Hamiltonian $H_{ZX}(T)$ may be that it is composed of all-to-all interactions, resulting in a large norm of the total Hamiltonian such as $|\!| H_{ZX}(T) |\!|_{\infty} = O(N^2)$ for any $T \geq 0$. 
%When one compares the cost of this Hamiltonian dynamics with that of parallel quantum circuits, it may not be fair to use the Hamiltonian with such a large norm. 
%However, we are here more interested in the implementations of the dynamics in physically feasible many-body systems. 
%The $H_{ZX}(T)$ is especially meaningful from this point of view because the Hamiltonians consisting $H_{ZX}(T)$ are indeed common in e.g. the cavity QED, where all-to-all interactions naturally appear due to the cavity modes. 
%In this context, it is also more important to consider the norm of each \emph{local term}, not that of the total Hamiltonian, as it describes the strength of each interaction.
%Since the norm of each local term of the proposed Hamiltonian $H_{XZ}(T)$ is less than one and is properly normalised, we may fairly conclude that that the Hamiltonian actually generates a unitary $2$-design in a constant time.

In the study of random process in complex many-body systems~\cite{PSW2006,GLTZ2006,R2008,HP2007,SS2008,LSHOH2013,SS2014,RD2015,SS2015}, it is important to fully clarify the microscopic dynamics leading to realisations of unitary designs in the systems, otherwise a solid basis of those studies remains lacking.
Although $H_{ZX}(T)$ is time-dependent and is composed of all-to-all interactions, Corollary~\ref{Cor:RHD} provides a relatively natural many-body Hamiltonian by which unitary $2$-designs can be generated. We believe that our result leads to new insights toward the full understandings of fundamental phenomena from the microscopic point of view.

\section{Proof of the main result} \label{Sect:Proofs}

A proof of Theorem~\ref{Thm:Diag2} is given in this section. To make the outline of the proof clear, we provide an overview in Subsection~\ref{SS:OP}. 
The key lemma for the proof is given in Subsection~\ref{SS:al} with its proof in Subsection~\ref{SS:LR}. The proof of Theorem~\ref{Thm:Diag2} is then given in Subsection~\ref{SS:CA}.

\subsection{Overview of the proof} \label{SS:OP}

In order to prove Theorem~\ref{Thm:Diag2}, we need to investigate $|\!| \mc{G}_{U[\ell]}^{(2)} - \mc{G}_{\sf H}^{(2)}|\!|_{\diamond}$.
As $\mc{G}_{\sf H}^{(2)}(\rho)$ for any $\rho \in \mc{B}(\mc{H}^{\otimes 2})$ can be explicitly obtained from the Schur-Weyl duality~\cite{GR1999}, the main task is to analyse $\mc{G}_{U[\ell]}^{(2)}$, which is reduced to the investigation of a map $\mc{R}=\mathcal{G}_{U^Z}^{(2)} \circ \mathcal{G}_{U^X}^{(2)} \circ \mathcal{G}_{U^Z}^{(2)}$ due to the independence of the random diagonal unitaries (see Eq.~\eqref{Eq:un}).

The proof of the achievability simply follows from the key lemma, which states that
\begin{equation}
\mc{R}^{\ell} =\bigl(1-\Theta(d^{-\ell}) \bigr) \mc{G}_{\sf H}^{(2)} +\Theta(d^{-\ell}) \mc{C}^{(\ell)},
\end{equation}
where $\mc{C}^{(\ell)}$ is a CPTP map dependent on $\ell$ (see Lemma~\ref{Lemma:R} in Subsection~\ref{SS:al}), making the intuition that $\mc{R}$ has a strong randomisation ability rigorous.
Hence, repetitions of $U'^{Z} U^X U^Z$ can be arbitrarily close to a Haar random unitary up to the second order. The complete argument is given in Subsection~\ref{SS:CA}.

To obtain the converse, we investigate a special case in detail, which is also presented in Subsection~\ref{SS:CA}. Together with the achievability and the converse, we obtain our main result that
\begin{equation}
|\!| \mc{G}_{U[\ell]}^{(2)} - \mc{G}_{\sf H}^{(2)}|\!|_{\diamond}
=\Theta(1/d^{\ell}),
\end{equation}
which implies that $U[\ell]$ is a $\Theta(d^{-\ell})$-approximate unitary $2$-design.

\subsection{Auxiliary lemmas} \label{SS:al}

Before we prove Theorem~\ref{Thm:Diag2}, we introduce additional notation and useful lemmas.
In the rest of the paper, the Pauli-$Z$ and -$X$ bases are always denoted by $\{ \ket{i} \}_{i=0, \cdots, d-1}$ (Latin letters) and $\{ \ket{\alpha} \}_{\alpha=0, \cdots, d-1}$ (Greek letters), respectively. 
We also define the entangled states $|\phi_{ij}^{(\pm)} \rangle:=\frac{1}{\sqrt{2}} (\ket{ij} \pm \ket{ji})$ for $i>j$.

We use several operators in $\mc{B}(\mc{H}^{\otimes 2})$. First, we denote by $\mathbb{I}$, $\mathbb{F}$ the identity operator and the swap operator defined by $\sum_{i,j} \ketbra{ij}{ji}$, respectively. 
We also define the basis dependent operators $\mathbb{L}^{(0)} := \sum_{i} \ketbra{ii}{ii}$ and
$\mathbb{L}^{(1)} := \sum_{i>j} |\phi_{ij}^{(+)} \rangle \langle \phi_{ij}^{(+)} |$.
We denote by $P^{\rm sym}$ and $P^{\rm anti}$ 
the projection operators onto the symmetric and antisymmetric subspaces of $\mc{H}^{\otimes 2}$ , which are equal to $(\mathbb{I}+\mathbb{F})/2$ and $(\mathbb{I}-\mathbb{F})/2$, respectively.
Note that $P^{\rm sym} = \mathbb{L}^{(0)}+\mathbb{L}^{(1)}$
and $P^{\rm anti} = \sum_{i>j} | \phi_{ij}^{(-)} \rangle \langle \phi_{ij}^{(-)}|$.
By normalising these operators, we define the following states: 
\begin{equation}
\Pi^{\rm sym}=2 P^{\rm sym}/d(d+1), \hspace{4mm} 
\Pi^{\rm anti} = 2 P^{\rm anti}/d(d-1), \hspace{4mm}
\Lambda^{(0)} =\mathbb{L}^{(0)}/d,\hspace{4mm}
\Lambda^{(1)}= \mathbb{L}^{(1)}/d.
\end{equation}

Throughout the proof, we denote the coefficients of $\ket{\alpha}$ in the basis of $\{ \ket{i} \}$ by $ \alpha_i/\sqrt{d}$, i.e. $\alpha_i = \sqrt{d} \braket{i}{\alpha}$. Similarly, we define $i_{\alpha}:= \sqrt{d} \braket{\alpha}{i}$. From the properties of the Pauli-$Z$ and -$X$ bases, it follows that $\alpha_i = i_{\alpha} \in \{\pm 1\}$.
We also define $f^{ij}_{kl}$ given by
\begin{equation}
f^{ij}_{kl} = \frac{2}{d^3}\biggl( \sum_{\alpha=0}^{d-1} \alpha_i \alpha_j \alpha_k \alpha_l  \biggr)^2,
\end{equation}
which satisfies the following properties (see~\ref{App:1} for the proof).

\begin{lemma} \label{Lemma:f}
{\it
The quantity $f^{ij}_{kl}$ is either $0$ or $2/d$, and  
satisfies $f^{ij}_{kl} = f^{kl}_{ij}$, $\sum_{i>j} f^{ij}_{kl}=1$ and
$\sum_{s>t} f^{ij}_{st} f^{st}_{kl}=f^{ij}_{kl}$.
}
\end{lemma}

Using $f^{ij}_{kl}$, we obtain the key lemma about the map $\mc{R}=\mathcal{G}_{U^Z}^{(2)} \circ \mathcal{G}_{U^X}^{(2)} \circ \mathcal{G}_{U^Z}^{(2)}$.

\begin{lemma} \label{Lemma:R}
Let $\ell$ be a natural number. Then, $\ell$ repetitions of the CPTP map $\mc{R}$, denoted by $\mc{R}^{\ell}$, is given by
\begin{equation}
\mc{R}^{\ell} = (1-p_{\ell}) \mc{G}_{\sf H}^{(2)} + p_{\ell} \mc{C}^{(\ell)},
\end{equation}
where $p_{\ell}=\frac{d^{\ell+1}+d^{\ell}-2}{d^{2\ell}(d-1)}$.
Here, $\mc{C}^{(\ell)}$ is a unital CPTP map given by
\begin{multline}
\mc{C}^{(\ell)}(\rho) =
\frac{1}{d(d^{\ell+1}+d^{\ell}-2)}
\biggl(
\bigl( (2d^{\ell}+d-3) \rho_0 + 2(d^{\ell}-1) \rho_1 \bigr) \Lambda^{(0)}
+
(d-1)(d^{\ell}-1) \rho_0 \Lambda^{(1)}\\
+
 2(d^{\ell}-1) \rho_2 \Pi^{\rm anti}
+
d^{\ell+1}(d-1)\sum_{i>j} \sum_{k>l} f^{ij}_{kl} \sum_{a=\pm} \langle\phi^{(a)}_{ij}| \rho |\phi^{(a)}_{ij}\rangle |\phi^{(a)}_{kl}\rangle \langle\phi^{(a)}_{kl}|
\biggr).
\end{multline}
where $\rho_0 = \tr \rho \mathbb{L}^{(0)}$, $\rho_1 = \tr \rho \mathbb{L}^{(1)}$, and $\rho_2 = \tr \rho P^{\rm anti}$.
\end{lemma}

As this is the main technical result, we prove it in the next section before we present the proof of our main result.

\subsection{Proof of Lemma~\ref{Lemma:R}} \label{SS:LR}

To show Lemma~\ref{Lemma:R}, we start with the following lemma, whose proof is given in~\ref{App:2}.

\begin{lemma} \label{Lemma:RR}
{\it
Let $B$ be the basis in $\mc{H}^{\otimes 2}$ given by $\{\ket{ii}\}_{i=0}^{d-1} \cup \{|\phi_{ij}^{(+)} \rangle\}_{i>j} \cup \{|\phi_{ij}^{(-)} \rangle\}_{i>j}$.
Then, $\mc{R}^\ell(\ketbra{p}{q}) = 0$ for all $\ket{p} \neq \ket{q}  \in B$ and all positive integers $\ell$, and
\begin{align}
&\mc{R}^{\ell}(\ketbra{ii}{ii}) = (1-d^{-2\ell}) \Pi^{\rm sym} + d^{-2\ell} \Lambda \\
&\mc{R}^{\ell}(| \phi_{i j}^{(+)} \rangle \langle \phi_{i j}^{(+)}|) =
(1-p_\ell) \Pi^{\rm sym} + q_\ell \Lambda
+d^{-\ell} \sum_{k>l}f^{i j}_{k l} |\phi_{kl}^{(+)} \rangle \langle \phi_{kl}^{(+)} | \\
&\mc{R}^\ell(| \phi_{i j}^{(-)} \rangle \langle \phi_{i j}^{(-)}|) =
(1-d^{-\ell}) \Pi^{\rm anti} 
+d^{-\ell}\sum_{ k>l} f^{ij}_{kl} |\phi_{kl}^{(-)} \rangle \langle \phi_{kl}^{(-)} |, \label{Eq:q34mtfavsd}
\end{align}
where $p_\ell =\frac{d^{\ell+1} + d^\ell - 2}{d^{2\ell} (d-1)}$ and $q_\ell= 2 \frac{d^\ell-1}{d^{2\ell} (d-1)}$.
}
\end{lemma}

Using Lemma~\ref{Lemma:RR}, we have that for all $\rho \in \mc{B}(\mc{H}^{\otimes 2})$,
\begin{multline}
\mc{R}^\ell(\rho)=
\bigl( (1-d^{-2\ell})\rho_0 + (1-p_\ell) \rho_1 \bigr) \Pi^{\rm sym} 
+ 
(d^{-2\ell} \rho_0 + q_\ell \rho_1) \Lambda 
+
(1-d^{-\ell})\rho_2 \Pi^{\rm anti} \\
+
d^{-\ell} \sum_{ i >j }\sum_{ k >l }f^{ij}_{kl} \sum_{a=\pm}
\langle \phi_{ij}^{(a)} | \rho | \phi_{ij}^{(a)} \rangle |\phi_{kl}^{(a)} \rangle \langle \phi_{kl}^{(a)} |,
\end{multline}
where $\rho_0=\tr \rho \mathbb{L}^{(0)}$, $\rho_1= \tr \rho \mathbb{L}^{(1)}$, and $\rho_2=\tr \rho P^{\rm anti}$.
On the other hand, due to the Schur-Weyl duality~\cite{GR1999}, it follows that 
\begin{align}
\mc{G}_{\sf H}^{(2)}(\rho) &= (\tr P^{\rm sym} \rho) \Pi^{\rm sym} + (\tr P^{\rm anti} \rho) \Pi^{\rm anti}\\
&=(\rho_0 + \rho_1) \Pi^{\rm sym} + \rho_2 \Pi^{\rm anti},
\end{align}
where we used a simple fact that $P^{\rm sym}=\mathbb{L}^{(0)} + \mathbb{L}^{(1)}$.
Thus, after some calculation, we obtain
\begin{equation}
\mc{R}^{\ell}(\rho) = (1-p_{\ell}) \mc{G}_{\sf H}^{(2)}(\rho) + p_{\ell} \mc{C}^{(\ell)}(\rho), \label{Eq:::::}
\end{equation}
where $\mc{C}^{(\ell)}$ is a linear map given by
\begin{multline}
\mc{C}^{(\ell)}(\rho) =
\frac{1}{d^{\ell+1}+d^{\ell}-2}
\biggl(
\bigl( (2d^{\ell}+d-3) \rho_0 + 2(d^{\ell}-1) \rho_1 \bigr) \Lambda^{(0)}
+
(d-1)(d^{\ell}-1)\rho_0 \Lambda^{(1)}\\
+
2(d^{\ell}-1) \rho_2 \Pi^{\rm anti}
+
d^{\ell }(d-1)\sum_{i>j} \sum_{k>l} f^{ij}_{kl} \sum_{a=\pm} \langle\phi^{(a)}_{ij}| \rho |\phi^{(a)}_{ij}\rangle |\phi^{(a)}_{kl}\rangle \langle\phi^{(a)}_{kl}|
\biggr). \label{Eq:2m3ko2143af}
\end{multline}
In the following, we show that $\mc{C}^{(\ell)}$ is a unital CPTP map.

The unitality can be easily observed from Eq.~\eqref{Eq:::::} because both $\mc{R}^{\ell}$ and $\mc{G}_{\sf H}^{(2)}$ are averages of unitary conjugations and hence are unital.
To show the complete positivity and the trace preserving property, we use the Choi-Jamio\l kowski representation of a linear map~\cite{J1972,C1975}, which is an isomorphism $J$ between a set of linear operators $\mc{E}: \mc{B}(\mc{H}) \rightarrow \mc{B}(\mc{H'})$ and a set of operators $\mc{B}(\mc{H} \otimes \mc{H'})$
\begin{equation}
J(\mc{E}) = (\mc{E} \otimes {\rm id}_{|\mc{H}|})(\ketbra{\Phi}{\Phi}),
\end{equation}
where ${\rm id}_{|\mc{H}|}$ is the identity map on $\mc{H}$, and $\ket{\Phi}=\frac{1}{\sqrt{|\mc{H}|}}\sum_i \ket{ii}$ is the maximally entangled state.
The Choi-Jamio\l kowski representation has properties that 
$\mc{E}$ is CP if and only if $J(\mc{E}) \geq 0$,
and that $\mc{E}$ is TP if and only if $\tr_{\mc{H}'} J(\mc{E}) = I_\mc{H}/|\mc{H}|$~\cite{W2012}.

To obtain the Choi-Jamio\l kowski representation of $\mc{C}^{(\ell)}$, it is convenient to use the basis $B =\{\ket{ii}\}_{i=0}^{d-1} \cup \{|\phi_{ij}^{(+)} \rangle\}_{i>j} \cup \{|\phi_{ij}^{(-)} \rangle\}_{i>j}$. Then, we have
\begin{multline}
J(\mc{C}^{(\ell)}) =
\frac{1}{d^2} 
\sum_{i} \mc{C}^{(\ell)}(\ketbra{ii}{jj}) \otimes \ketbra{ii}{jj}
+
\frac{1}{d^2} 
\sum_{i>j} \sum_{k>l} \sum_{a=\pm} \mc{C}^{(\ell)}(|\phi_{ij}^{(a)}\rangle \langle \phi_{kl}^{(a)}|) \otimes |\phi_{ij}^{(a)}\rangle \langle \phi_{kl}^{(a)}|.
\end{multline}
As an explicit form of $\mc{C}^{(\ell)}$ is given in Eq.~\eqref{Eq:2m3ko2143af},
we obtain
\begin{multline}
J(\mc{C}^{(\ell)}) =
\frac{1}{d(d^{\ell+1}+d^{\ell}-2)}
\bigl(
(d-1)(d^{\ell}-1)(\Lambda^{(0)} \otimes \Lambda^{(1)} + \Lambda^{(1)} \otimes \Lambda^{(0)} + \Pi^{\rm anti} \otimes \Pi^{\rm anti})\\
+
(2d^{\ell}+d-3)\Lambda^{(0)} \otimes \Lambda^{(0)}
+
d^{\ell-1}(d-1)\sum_{i>j} \sum_{k>l} f^{ij}_{kl} \sum_{a=\pm} |\phi^{(a)}_{kl}\rangle \langle\phi^{(a)}_{kl}| \otimes |\phi^{(a)}_{ij} \rangle \langle \phi^{(a)}_{ij}|
\bigr).
\end{multline}
Noting that this is already diagonal in the basis of $B$ and all the coefficients are non-negative, since $f^{ij}_{kl} \in \{0, 2/d \}$ from Lemma~\ref{Lemma:f}, it is obvious that $J(\mc{C}^{(\ell)}) \geq 0$, implying that $\mc{C}^{(\ell)}$ is a CP map.
It is also straightforward to check $\tr_{\mc{H}'} J(\mc{E}) = I_\mc{H}/|\mc{H}|$ by using a relation $\sum_{k>l} f^{ij}_{kl} = 1$ for any $i>j$, and hence $\mc{C}^{(\ell)}$ is a TP map. $\hfill \blacksquare$

\subsection{Proof of the main result} \label{SS:CA}

We now prove Theorem~\ref{Thm:Diag2}.
We first show that $|\!| \mc{G}_{U[\ell]}^{(2)} - \mc{G}_{\sf H}^{(2)}|\!|_{\diamond}=O(1/d^{\ell})$.
The $\mc{G}_{U[\ell]}^{(2)}$ is equal to $\mc{R}^\ell$, where $\mc{R}=\mathcal{G}_{U^Z}^{(2)} \circ \mathcal{G}_{U^X}^{(2)} \circ \mathcal{G}_{U^Z}^{(2)}$, since for any $\rho \in \mc{B}(\mc{H})$
\begin{align}
\mc{G}_{U[\ell]}^{(2)}(\rho) &= \mathbb{E}_{U[\ell]} [ (U[\ell])^{\otimes 2} \rho (U[\ell])^{\dagger \otimes 2}]\\
&= \prod_{i=1}^\ell  \mathbb{E}_{U_i^{'Z}} \mathbb{E}_{U_i^X} \mathbb{E}_{U_i^Z} [ \bigl(U_i^{'Z} U_i^X U_i^Z \bigr)^{\otimes 2} \rho \bigl(U_i^{'Z} U_i^X U_i^Z \bigr)^{\dagger \otimes 2}]\\
&= \bigl(\mathcal{G}_{U^Z}^{(2)} \circ \mathcal{G}_{U^X}^{(2)} \circ \mathcal{G}_{U^X}^{(2)} \bigr)^\ell(\rho)\\
&=\mc{R}^\ell(\rho),
\end{align}
where the second line is obtained using the fact that each random diagonal unitary is independent.
Hence, $|\!| \mc{G}_{U[\ell]}^{(2)} - \mc{G}_{\sf H}^{(2)}|\!|_{\diamond}=|\!| \mc{R}^\ell- \mc{G}_{\sf H}^{(2)}|\!|_{\diamond}$.
From Lemma~\ref{Lemma:R}, an upper bound is now easily obtained:
\begin{align}
|\!|\mc{R}^\ell - \mc{G}_{\sf H}^{(2)}|\!|_{\diamond}
&= 
|\!| p_{\ell} \bigl(\mc{C}^{(\ell)}-\mc{G}_{\sf H}^{(2)} \bigr)|\!|_{\diamond}\\
&\leq
p_{\ell} \bigl(|\!|\mc{C}^{(\ell)}|\!|_{\diamond} + |\!|\mc{G}_{\sf H}^{(2)}|\!|_{\diamond} \bigr)\\
&\leq
2 p_{\ell}\\
&\leq
\frac{2}{d^\ell}\left[ 1+ \frac{2}{d-1} \right],
\end{align}
where we have used the triangle inequality in the second line and the fact that $\mc{C}^{(\ell)}$ and $\mc{G}_{\sf H}^{(2)}$ are CPTP maps in the third line.

To obtain the converse, i.e. $|\!|\mc{R}^\ell - \mc{G}_{\sf H}^{(2)}|\!|_{\diamond} = \Omega(1/d^{\ell})$, we use a fact that $|\!| \mc{R}^\ell(\rho) - \mc{G}_{\sf H}^{(2)}(\rho)|\!|_1 \leq |\!| \mc{R}^\ell - \mc{G}_{\sf H}^{(2)}|\!|_{\diamond}$ for any $\rho \in \mc{S}(\mc{H})$, which follows from the definition of the diamond norm. 
By substituting $| \phi_{i_0j_0}^{(+)} \rangle \langle \phi_{i_0j_0}^{(+)}|$ ($i_0>j_0$) and using Lemma~\ref{Lemma:R}, we obtain
\begin{align}
|\!| \mc{R}^\ell(| \phi_{i_0j_0}^{(+)} \rangle \langle \phi_{i_0j_0}^{(+)}|) - \mc{G}_{\sf H}^{(2)}(| \phi_{i_0j_0}^{(+)} \rangle \langle \phi_{i_0j_0}^{(+)}|)|\!|_1
&=p_\ell |\!| \mc{C}^{(\ell)}(| \phi_{i_0j_0}^{(+)} \rangle \langle \phi_{i_0j_0}^{(+)}|) - \mc{G}_{\sf H}^{(2)}(| \phi_{i_0j_0}^{(+)} \rangle \langle \phi_{i_0j_0}^{(+)}|)|\!|_1.
\end{align}
Now, the operator $\mc{R}^\ell(\Phi_{i_0j_0}) - \mc{G}_{\sf H}^{(2)}(\Phi_{i_0j_0})$ can be directly calculated to be
\begin{multline}
\mc{R}^\ell(\Phi_{i_0j_0}) - \mc{G}_{\sf H}^{(2)}(\Phi_{i_0j_0})
=
-2\frac{d-1}{(d+1)(d^{\ell+1} + d^\ell-2)} \Lambda^{(0)}\\
+
\sum_{k>l}\bigl( \frac{d^\ell(d-1)}{d^{\ell+1} + d^\ell-2} f^{i_0j_0}_{kl} - \frac{2}{d(d+1)} \bigr) |\phi_{kl}^{(+)} \rangle \langle \phi_{kl}^{(+)}|,
\end{multline}
which is already diagonal in the basis of $B =\{\ket{ii}\}_{i=0}^{d-1} \cup \{|\phi_{ij}^{(+)} \rangle\}_{i>j} \cup \{|\phi_{ij}^{(-)} \rangle\}_{i>j}$.
Since $f_{kl}^{i_0j_0}$ satisfies $f_{kl}^{i_0j_0} \in \{ 0, 2/d\}$ for any $k>l$ and $\sum_{k>l} f_{kl}^{i_0j_0} =1$ from Lemma~\ref{Lemma:f}, the number of $(k,l)$ ($k>l$) for which $f_{kl}^{i_0j_0}$ is nonzero is $d/2$.
Using this fact, we obtain
\begin{equation}
|\!| \mc{R}^\ell(\Phi_{i_0j_0}) - \mc{G}_{\sf H}^{(2)}(\Phi_{i_0j_0})|\!|_1
=
\frac{2}{d^{\ell}} 
                  - 2 \frac{d^{\ell+1} + d^\ell -2}{d^{2\ell} (d^2-1)}
            \geq \frac{2}{d^\ell}\left[ 1-\frac{1}{d-1}\right],
\end{equation}
providing a lower bound of $|\!|\mc{R}^\ell - \mc{G}_{\sf H}^{(2)}|\!|_{\diamond}$.

From these bounds and recalling that $\mc{R}^\ell =\mc{G}_{U[\ell]}^{(2)}$,
we obtain tight upper and lower bounds of $|\!|\mc{G}_{U[\ell]}^{(2)} - \mc{G}_{\sf H}^{(2)}|\!|_{\diamond}$ as
\begin{equation}
\frac{2}{d^\ell}\left[ 1-\frac{1}{d-1}\right] \leq
|\!| \mc{G}_{U[\ell]}^{(2)} - \mc{G}_{\sf H}^{(2)}|\!|_{\diamond}
\leq \frac{2}{d^\ell}\left[ 1+ \frac{2}{d-1}\right]. \label{Eq:;3qn4||}
\end{equation}
This implies that $U[\ell]$ is a $\Theta(d^{-\ell})$-approximate unitary 2-design and concludes the proof.
$\hfill \blacksquare$

\section{Conclusion}\label{Sect:Conclusion}

We have proven that an approximate unitary $2$-design can be achieved by alternately and repeatedly applying independent random $Z$- and $X$-diagonal unitaries. 
More specifically, we showed that it converges to an $\epsilon$-approximate unitary $2$-design exponentially quickly in terms of the number $\ell$ of repetitions, namely $\epsilon = \Theta(1/d^{\ell})$.
As it provides the converse, our result can be used to investigate the optimality of using unitary $2$-designs in quantum protocols (see e.g.~\cite{NHMW2015-2}).
Based on this result, we constructed an efficient quantum circuit composed of $\Theta\bigl(N(N+\log1/\epsilon)\bigr)$ gates that implement an $\epsilon$-approximate unitary $2$-design.
The merit of the construction is that most of the gates are diagonal in the Pauli- $Z$ basis and the non-commuting part is of constant depth, which may result in the quick and robust implementations of the design. Finally, by interpreting the quantum circuits in terms of Hamiltonian dynamics, we have shown that 
a time-dependent Hamiltonian consisting of disordered Hamiltonians generates unitary $2$-designs after switching the interactions a few times.

A natural question is whether or not more repetitions eventually achieve unitary $t$-designs for arbitrary $t$. This problem has been subsequently addressed by some of the present authors in Ref.~\cite{NHKW2016}.
It is also interesting to consider an approximate implementation of random $Z$-diagonal unitaries by quantum circuits where each gate acts only on neighboring qubits. If there exists such a quantum circuit, it immediately provides a random Hamiltonian with neighboring interactions that realises unitary $2$-designs, which have strong contributions to the microscopic understandings of fundamental physics resulted from random dynamics in complex systems.

%\section{Acknowledgements}
\begin{acknowledgments}
The authors are grateful to W. Brown, R. F. Werner, and O. Fawzi for interesting
and fruitful discussions. 
YN is supported by JSPS Postdoctoral Fellowships for Research Abroad and partially by JSPS KAKENHI Grant Number 272650.
CH and CM acknowledge support from the EU grants SIQS and QFTCMPS and by the cluster of excellence EXC 201 Quantum Engineering and Space-Time Research.
CH is further supported by FPI Grant No. BES-2014-068888.
AW is supported by the European Commission (STREP ``RAQUEL''), the European Research Council (Advanced Grant ``IRQUAT''), the Spanish MINECO, project FIS2008-01236, with the support of FEDER funds.
CH and AW are also supported by the Generalitat de Catalunya, CIRIT project no. 2014 SGR 966,
as well as the Spanish MINECO, projects FIS2013-40627-P and FIS2016-80681-P (AEI/FEDER, UE)
\end{acknowledgments}

\appendix

\section{Proof of Lemma~\ref{Lemma:f}} \label{App:1}

A proof for a technical lemma~\ref{Lemma:f} is provided. The statement is about the properties of  $f^{ij}_{kl}$ given by
\begin{equation}
f^{ij}_{kl} = \frac{2}{d^3}\biggl( \sum_{\alpha=0}^{d-1} \alpha_i \alpha_j \alpha_k \alpha_l  \biggr)^2,
\end{equation}
and we show that $f^{ij}_{kl}$ is in $\{ 0, 2/d \}$, $f^{ij}_{kl} = f^{kl}_{ij}$, $\sum_{i>j} f^{ij}_{kl}=1$, and $\sum_{s>t} f^{ij}_{st} f^{st}_{kl}=f^{ij}_{kl}$.

\begin{Proof}[Lemma~\ref{Lemma:f}]
The statement $f^{ij}_{kl} = f^{kl}_{ij}$ follows from the definition of $f^{ij}_{kl}$.
We first show that $f^{ij}_{kl}$ is either $0$ or $2/d$.
As $f^{ij}_{kl}$ is defined by $f^{ij}_{kl} = \frac{2}{d^3} \bigl( \sum_{\alpha=0}^{d-1}  \alpha_i\alpha_j\alpha_k\alpha_l  \bigr)^2$,
we investigate $\sum_{\alpha=0}^{d-1}  \alpha_i\alpha_j\alpha_k\alpha_l$.
This is invariant even if Pauli $X$ is applied on the $m$-th qubit for any $m \in [1, \cdots, N]$, which we denote by $X_m$, since
\begin{align}
\sum_{\alpha=0}^{d-1}  \alpha_i\alpha_j\alpha_k\alpha_l &= d^2  \sum_{\alpha=0}^{d-1}  \braket{\alpha}{i} \braket{\alpha}{j}\braket{\alpha}{k}\braket{\alpha}{l}\\
&=d^2  \sum_{\alpha=0}^{d-1}  \bra{\alpha}X_m \ket{i} \bra{\alpha}X_m \ket{j}\bra{\alpha}X_m \ket{k}\bra{\alpha}X_m \ket{l}.
\end{align}
This is due to $\bra{\alpha}X_m = \pm \bra{\alpha}$. Hence, we assume $\ket{i} = \ket{0}^{\otimes N}$ without loss of generality, resulting in $\alpha_i=1$ for all $\alpha$.
The $\sum_{\alpha=0}^{d-1}  \alpha_j\alpha_k\alpha_l$ has yet another invariance, that is,
\begin{align}
\sum_{\alpha=0}^{d-1}  \alpha_j\alpha_k\alpha_l &= d \sqrt{d}  \sum_{\alpha=0}^{d-1} \braket{\alpha}{j}\braket{\alpha}{k}\braket{\alpha}{l}\\
&=d \sqrt{d}  \sum_{\alpha=0}^{d-1}  \bra{\alpha}Z_m \ket{j}\bra{\alpha}Z_m \ket{k}\bra{\alpha}Z_m \ket{l},
\end{align}
due to the summation over all $\alpha$, where $Z_m$ is the Pauli-$Z$ operator acting on the $m$-th qubit.
We then assume $\alpha_j = 1$ for $j=0,\cdots, d/2-1$ and $\alpha_j = -1$ for $j=d/2,\cdots, d-1$ without loss of generality.
This leads to 
\begin{equation}
\sum_{\alpha=0}^{d-1}  \alpha_i\alpha_j\alpha_k\alpha_l 
=\biggl( \sum_{\alpha=0}^{d/2-1} - \sum_{\alpha=d/2}^{d-1} \biggr) \alpha_k\alpha_l.
\end{equation}
Denoting $\ket{\alpha}$ by $\ket{\alpha^1 \alpha^2 \cdots \alpha^N}$ ($\alpha^m = \pm$), where $\ket{\pm}$ are the eigenbasis of the Pauli-$X$ with eigenvalues $\pm1$, respectively, and similarly denoting $\ket{k}$ and $\ket{l}$ in binary such as $\ket{k^1 \cdots k^N}$ ($k_m = 0,1$),
$(\sum_{\alpha=0}^{d/2-1} - \sum_{\alpha=d/2}^{d-1} ) \alpha_k \alpha_l $ is rewritten as
\begin{multline}
d \times \sum_{\alpha^2, \cdots, \alpha^N = \pm} \biggl( \braket{+}{k^1} \braket{+}{l^1} \braket{\alpha^2 \cdots \alpha^N}{k^1 \cdots k^N} \braket{\alpha^2 \cdots \alpha^N}{l^1 \cdots l^N}\\
-
\braket{-}{k^1} \braket{-}{l^1}
\braket{\alpha^2 \cdots \alpha^N}{k^1 \cdots k^N} \braket{\alpha^2 \cdots \alpha_N}{l^1 \cdots l^N} \biggr).
\end{multline}
When $k^1=l^1$, this is zero. When $k^1 \neq l^1$, this is equal to $d \prod_{m=2}^N \delta_{k^m,l^m}$, which is either $0$ or $d$. Thus, $f_{ij}^{kl} \in \{ 0, 2/d \}$. 

We next show $\sum_{k>l} f^{ij}_{kl}=1$ for any $i>j$.
\begin{align}
\sum_{k>l} f^{ij}_{kl} &=\frac{2}{d^3} \sum_{k>l} \biggl( \sum_{\alpha} \alpha_i \alpha_j\alpha_k\alpha_l\biggr)^2\\
&=\frac{1}{d^3}  \sum_{\alpha, \beta}  \alpha_i \alpha_j  \beta_i \beta_j  \biggl( \sum_{k,l}\alpha_k\alpha_l\beta_k\beta_l - \sum_{k}\alpha_k^2\beta_k^2 \biggr).
\end{align}
As $\sum_{k}\alpha_k^2\beta_k^2 = d$ due to $\alpha_k, \beta_k \in \{\pm1\}$, the later term is given by 
\begin{align}
\frac{1}{d^3}  \sum_{\alpha, \beta}  \alpha_i \alpha_j  \beta_i \beta_j   \sum_{k}\alpha_k^2\beta_k^2
&=\frac{1}{d^2}  \sum_{\alpha, \beta}  \alpha_i \alpha_j  \beta_i \beta_j  \\
&= \biggl( \sum_{\alpha}  \braket{i}{\alpha} \braket{\alpha}{j} \biggr)^2\\
&=0,
\end{align}
where we used that $i \neq j$ for the last line. Hence,
\begin{align}
\sum_{k>l} f^{ij}_{kl} =\frac{1}{d^3}  \sum_{\alpha, \beta}  \alpha_i \alpha_j  \beta_i \beta_j  \biggl( \sum_{k}\alpha_k\beta_k \biggr)^2.
\end{align}
As $\sum_{k}\alpha_k\beta_k $ is given by $\frac{1}{d^2}\sum_k \braket{\alpha}{k}\braket{k}{\beta}=\frac{1}{d^2}\delta_{\alpha \beta}$, we obtain
\begin{equation}
\sum_{k>l} f^{ij}_{kl} =\frac{1}{d}  \sum_{\alpha, \beta}  \alpha_i \alpha_j  \beta_i \beta_j  \delta_{\alpha,\beta}=1.
\end{equation}

We finally show $\sum_{s>t} f^{ij}_{st} f^{st}_{kl}=f^{ij}_{kl}$.
To this end, we define a set $\Xi_{ij}$ for $i>j$ by
$\Xi_{ij} := \bigl\{ (s,t) | s,t \in \{1,\cdots, N \}, s>t, f_{st}^{ij} = \frac{2}{d}  \bigr\}$.
Since $f_{kl}^{ij} \in \{ 0, 2/d \}$ and $\sum_{k>l} f_{kl}^{ij}=1$ for any $i>j$, the number of elements in $\Xi_{ij}$,denoted by $| \Xi_{ij} |$, is $d/2$.
Due to the definition of $f_{st}^{ij}$, $\Xi_{ij}$ is also given in terms of $\alpha_i$'s by
$\Xi_{ij} = \bigl\{ (s,t) | s,t \in \{1,\cdots, N \}, s>t, \forall \alpha \in [0, \cdots, d-1], \alpha_s \alpha_t = \alpha_i \alpha_j \bigr\}$.
From this, it is observed that $\forall i>j$ and $\forall k>l$, $\Xi_{ij}$ is either equal to $\Xi_{kl}$ or has no intersection with $\Xi_{kl}$, i.e. $\Xi_{ij} \cap \Xi_{kl} = \emptyset$.

In terms of $\Xi_{ij}$, $f_{ij}^{kl}=\frac{2}{d} \delta_{kl \in  \Xi_{ij}}$, where $\delta_{kl \in  \Xi_{ij}} = 1$ if $(k,l) \in  \Xi_{ij}$ and $0$ otherwise. Note that, as $f_{ij}^{kl}=f^{ij}_{kl}$, $\delta_{kl \in  \Xi_{ij}}=\delta_{ij \in  \Xi_{kl}}$. Using this notation, we have
\begin{align}
\sum_{s>t} f_{st}^{ij} f_{kl}^{st} &= \biggl( \frac{2}{d} \biggr)^2 \sum_{s>t} \delta_{st \in  \Xi_{kl}}\delta_{st \in  \Xi_{ij}}\\
&=
\biggl( \frac{2}{d} \biggr)^2 \sum_{s>t} \delta_{st \in  \Xi_{kl} \cap  \Xi_{ij}}. \label{Eq:last}
\end{align}
When $\Xi_{kl} =  \Xi_{ij}$, this is equal to $\frac{2}{d}$ as $|\Xi_{kl}|=d/2$.
In this case, $ f_{ij}^{kl} =\frac{2}{d} \delta_{kl \in  \Xi_{ij}}= \frac{2}{d}$ since $(k,l) \in \Xi_{kl} = \Xi_{ij}$, implying
$\sum_{s>t} f_{st}^{ij} f_{kl}^{st} = f_{ij}^{kl}$.
When $\Xi_{kl} \cap  \Xi_{ij}=\emptyset$, Eq.~\eqref{Eq:last} is equal to zero, and $f_{ij}^{kl}$ is also zero by definition. Hence, $\sum_{s>t} f_{st}^{ij} f_{kl}^{st} = f_{ij}^{kl}$ holds even in this case.
Since $\Xi_{ij}$ is either $\Xi_{kl}$ or satisfies $\Xi_{ij} \cap \Xi_{kl} = \emptyset$, this concludes the proof. $\hfill \blacksquare$
\end{Proof}

\section{Proof of Lemma~\ref{Lemma:RR}} \label{App:2}

We now prove the key technical lemma about the map $\mc{R}=\mathcal{G}_{U^Z}^{(2)} \circ \mathcal{G}_{U^X}^{(2)} \circ \mathcal{G}_{U^Z}^{(2)}$. The statement is as follows:
let $B$ be the basis in $\mc{H}^{\otimes 2}$ given by $\{\ket{ii}\}_{i=0}^{d-1} \cup \{|\phi_{ij}^{(+)} \rangle\}_{i>j} \cup \{|\phi_{ij}^{(-)} \rangle\}_{i>j}$, where
$|\phi_{ij}^{(\pm)} \rangle:=\frac{1}{\sqrt{2}} (\ket{ij} \pm \ket{ji})$.
Then, $\mc{R}^\ell(\ketbra{p}{q}) = 0$ for all $\ket{p} \neq \ket{q}  \in B$ and all integers $\ell$, and
\begin{align}
&\mc{R}^{\ell}(\ketbra{ii}{ii}) = (1-d^{-2\ell}) \Pi^{\rm sym} + d^{-2\ell} \Lambda \\
&\mc{R}^{\ell}(| \phi_{i j}^{(+)} \rangle \langle \phi_{i j}^{(+)}|) =
(1-p_\ell) \Pi^{\rm sym} + q_\ell \Lambda
+d^{-\ell} \sum_{k>l}f^{i j}_{k l} |\phi_{kl}^{(+)} \rangle \langle \phi_{kl}^{(+)} | \\
&\mc{R}^\ell(| \phi_{i j}^{(-)} \rangle \langle \phi_{i j}^{(-)}|) =
(1-d^{-\ell}) \Pi^{\rm anti} 
+d^{-\ell}\sum_{ k>l} f^{ij}_{kl} |\phi_{kl}^{(-)} \rangle \langle \phi_{kl}^{(-)} |, 
\end{align}
where $p_\ell =\frac{d^{\ell+1} + d^\ell - 2}{d^{2\ell} (d-1)}$ and $q_\ell= 2 \frac{d^\ell-1}{d^{2\ell} (d-1)}$.

\begin{Proof}[Lemma~\ref{Lemma:RR}]
We first investigate $\mc{R}(\ketbra{ii}{kk})$, 
$\mc{R}(|\phi_{ij}^{(+)} \rangle \langle \phi_{kl}^{(+)} |)$, and $\mc{R}(|\phi_{ij}^{(-)} \rangle \langle \phi_{kl}^{(-)} |)$ 
($i>j$ and $k>l$).
As each input state is in the Pauli-$Z$ basis, we obtain
\begin{align}
&\mc{R}(\ketbra{ii}{kk}) = \delta_{ik} \mathcal{G}_{U^Z}^{(2)} \circ \mathcal{G}_{U^X}^{(2)}(\ketbra{ii}{ii})\\
&\mc{R}(|\phi_{ij}^{(+)} \rangle \langle \phi_{kl}^{(+)} |) = \delta_{ik} \delta_{jl} \mathcal{G}_{U^Z}^{(2)} \circ \mathcal{G}_{U^X}^{(2)}(|\phi_{ij}^{(+)} \rangle \langle \phi_{ij}^{(+)} |) \\
&\mc{R}(|\phi_{ij}^{(-)} \rangle \langle \phi_{kl}^{(-)} |) = \delta_{ik} \delta_{jl} \mathcal{G}_{U^Z}^{(2)} \circ \mathcal{G}_{U^X}^{(2)}(|\phi_{ij}^{(-)} \rangle \langle \phi_{ij}^{(-)} |).
\end{align}
Using the relation $\mathcal{G}_{U^X}^{(2)}(\ketbra{ii}{ii}) = \frac{1}{d^2}\bigl( \mathbb{I} + \mathbb{F} - \mathbb{L}_X \bigr)$, where $\mathbb{L}_X = \sum_{\alpha} \ketbra{\alpha \alpha}{\alpha \alpha}$, and $\mathbb{I}$ and $\mathbb{F}$ are invariant under $\mathcal{G}_{U^Z}^{(2)}$,
the $\mc{R}(\ketbra{ii}{kk})$ is calculated to be 
\begin{equation}
\mc{R}(\ketbra{ii}{kk}) = \frac{1}{d^2} \delta_{ik}\biggl[ \bigl( 1- \frac{1}{d} \bigr)\bigl( \mathbb{I} + \mathbb{F} \bigr) + \frac{1}{d} \mathbb{L}  \biggr]. \label{Eq:ijkl}
\end{equation}
Note that this implies that $\mc{R}(\ketbra{ii}{ii})$ is independent of $i$.
For $\mc{R}(|\phi_{ij}^{(+)} \rangle \langle \phi_{kl}^{(+)} |) $ and $\mc{R}(|\phi_{ij}^{(-)} \rangle \langle \phi_{kl}^{(-)} |) $,
simple calculations lead to
\begin{align}
&\mathcal{G}_{U^X}^{(2)}(\ketbra{ij}{ij}) = \frac{1}{d^2}\biggl( \mathbb{I} + \sum_{\alpha, \beta} \alpha_i \alpha_j \beta_i \beta_j \ketbra{\alpha\beta}{\beta \alpha} - \mathbb{L}_X \biggr) \label{Eq:a324no}\\
&\mathcal{G}_{U^X}^{(2)}(\ketbra{ij}{ji}) = \frac{1}{d^2}\biggl(\sum_{\alpha, \beta} \alpha_i \alpha_j \beta_i \beta_j \ketbra{\alpha\beta}{\alpha \beta} +  \mathbb{F}  - \mathbb{L}_X \biggr) \label{Eq:b324no},
\end{align}
and similar relations for $\mathcal{G}_{U^Z}^{(2)}(\ketbra{\alpha \beta}{\alpha \beta})$ and $\mathcal{G}_{U^Z}^{(2)}(\ketbra{\alpha \beta}{\beta \alpha })$.
Hence, we obtain
\begin{align}
&\mc{R}(|\phi_{ij}^{(+)} \rangle \langle \phi_{kl}^{(+)} |) =\frac{1}{d^2} \delta_{ik} \delta_{jl} 
\biggl[
\bigl( 1- \frac{2}{d} \bigr)\bigl( \mathbb{I} + \mathbb{F} \bigr) + \frac{2}{d} \mathbb{L}
+
d \sum_{s>t} f^{ij}_{st}|\phi_{st}^{(+)} \rangle \langle \phi_{st}^{(+)} | \biggr] \label{Eq:R1symantisym0} \\
&\mc{R}(|\phi_{ij}^{(-)} \rangle \langle \phi_{kl}^{(-)} |) = 
\frac{1}{d^2}\delta_{ik} \delta_{jl} 
\biggl[
\mathbb{I} - \mathbb{F}
+
d \sum_{s>t} f^{ij}_{st}|\phi_{st}^{(-)} \rangle \langle \phi_{st}^{(-)} | \biggr], \label{Eq:R1symantisym}
\end{align}
where we use, e.g. $\alpha_i= i _{\alpha}$ for the derivation.

We next show that other terms, such as $\mc{R}(|\phi_{ij}^{(+)} \rangle \langle kk|)$, $\mc{R}(|\phi_{ij}^{(-)} \rangle \langle kk|)$, $\mc{R}(|\phi_{ij}^{(+)} \rangle \langle \phi_{kl}^{(-)} |)$ and their conjugates, are zero.
Amongst these terms, all except $\mc{R}(|\phi_{ij}^{(+)} \rangle \langle \phi_{ij}^{(-)} |)$ and its conjugate vanish after the first application of $\mathcal{G}_{U^Z}^{(2)}$.
For $\mc{R}(|\phi_{ij}^{(+)} \rangle \langle \phi_{ij}^{(-)} |)$,
$\mc{R}(|\phi_{ij}^{(+)} \rangle \langle \phi_{ij}^{(-)} |) = \mathcal{G}_{U^Z}^{(2)} \circ \mathcal{G}_{U^X}^{(2)}(|\phi_{ij}^{(+)} \rangle \langle \phi_{ij}^{(-)} |)$, since $|\phi_{ij}^{(+)} \rangle \langle \phi_{ij}^{(-)} |$ is not changed by $\mathcal{G}_{U^Z}^{(2)}$.
The $\mathcal{G}_{U^X}^{(2)}(|\phi_{ij}^{(+)} \rangle \langle \phi_{ij}^{(-)} |)$ term is expanded to be
\begin{align}
\mathcal{G}_{U^X}^{(2)}(|\phi_{ij}^{(+)} \rangle \langle \phi_{ij}^{(-)} |)
&=\frac{1}{2}\biggl(
\mathcal{G}_{U^X}^{(2)}(\ketbra{ij}{ij})
-\mathcal{G}_{U^X}^{(2)}(\ketbra{ij}{ji})
+\mathcal{G}_{U^X}^{(2)}(\ketbra{ji}{ij})
-\mathcal{G}_{U^X}^{(2)}(\ketbra{ji}{ji}).
\biggr)
\end{align}
This is calculated using Eqs.~\eqref{Eq:a324no} and~\eqref{Eq:b324no}. As the right hand sides of both Eqs.~\eqref{Eq:a324no} and~\eqref{Eq:b324no} are invariant under the exchange of $i$ and $j$, $\mathcal{G}_{U^X}^{(2)}(|\phi_{ij}^{(+)} \rangle \langle \phi_{ij}^{(-)} |)$ is zero, which implies $\mc{R}(|\phi_{ij}^{(+)} \rangle \langle \phi_{ij}^{(-)} |)=\mc{R}(|\phi_{ij}^{(-)} \rangle \langle \phi_{ij}^{(+)} |)=0$.

In the following, we investigate $\mc{R}^\ell(\ketbra{ii}{ii})$, $\mc{R}^\ell(|\phi_{ij}^{(+)} \rangle \langle \phi_{ij}^{(+)} |)$, and $\mc{R}^\ell(|\phi_{ij}^{(-)} \rangle \langle \phi_{ij}^{(-)} |)$.
Since we have  
\begin{equation}
\mc{R}(\mathbb{L}) = \frac{1}{d} \biggl[ \bigl( 1- \frac{1}{d} \bigr)\bigl( \mathbb{I} + \mathbb{F} \bigr) + \frac{1}{d} \mathbb{L}  \biggr],
\end{equation}
from Eq.~\eqref{Eq:ijkl}, $\mc{R}(\mathbb{I})=\mathbb{I}$, and $\mc{R}(\mathbb{F})=\mathbb{F}$,
it is observed from Eq.~\eqref{Eq:ijkl} that 
$\mc{R}^\ell(\ketbra{ii}{ii})$ is a linear combination of $\mathbb{I}+\mathbb{F}$ and $\mathbb{L}$.
Using this fact, it is straightforward to obtain
\begin{equation}
\mc{R}^\ell(\ketbra{ii}{ii}) = \frac{1- d^{-2\ell}}{d(d+1)} (\mathbb{I}+\mathbb{F})+ d^{-2\ell-1} \mathbb{L},
\end{equation}
which is rewritten, in terms of  $\Pi^{\rm sym} = \frac{1}{d(d+1)}(\mathbb{I} + \mathbb{F} )$ and $\Lambda=\frac{1}{d}\mathbb{L}$, as
\begin{equation}
\mc{R}^\ell(\ketbra{ii}{ii}) = (1- d^{-2\ell})\Pi^{\rm sym} + d^{-2\ell} \Lambda.
\end{equation}

Similarly, 
$\mc{R}^\ell(|\phi_{ij}^{(+)} \rangle \langle \phi_{ij}^{(+)} |)$ ($\mc{R}^\ell(|\phi_{ij}^{(-)} \rangle \langle \phi_{ij}^{(-)} |)$) is given by a linear combination of $\mathbb{I}+\mathbb{F}$, $\mathbb{L}$, and $ \sum_{s>t} f^{ij}_{st}|\phi_{st}^{(+)} \rangle \langle \phi_{st}^{(+)} |$
($\mathbb{I}- \mathbb{F}$ and $\sum_{s>t} f^{ij}_{st}|\phi_{st}^{(-)} \rangle \langle \phi_{st}^{(-)} |$).
This can be seen to hold, since
\begin{align}
\mc{R} \bigl(\sum_{s>t} f^{ij}_{st} |\phi_{st}^{(+)} \rangle \langle \phi_{st}^{(+)} | \bigr)
&=
\frac{1}{d^2}\biggl[ \bigl( 1- \frac{2}{d} \bigr)\bigl( \mathbb{I}+ \mathbb{F} \bigr) +\frac{2}{d} \mathbb{L} \biggr]
+
\frac{1}{d} \sum_{s>t} \sum_{k>l} f^{ij}_{st} f^{st}_{kl}|\phi_{kl}^{(+)} \rangle \langle \phi_{kl}^{(+)} |\\
&=
\frac{1}{d^2}\biggl[ \bigl( 1- \frac{2}{d} \bigr)\bigl( \mathbb{I}+ \mathbb{F} \bigr) +\frac{2}{d} \mathbb{L} \biggr]
+
\frac{1}{d} \sum_{k>l}f^{ij}_{kl} |\phi_{kl}^{(+)} \rangle \langle \phi_{kl}^{(+)} |, \label{Eq:fsphi}
\end{align}
where we have used $\sum_{s>t} f^{kl}_{st}=1$ and $\sum_{s>t} f^{ij}_{st} f^{st}_{kl}  = f^{ij}_{kl}$ due to Lemma~\ref{Lemma:f}, and similarly
\begin{align}
\mc{R} \bigl(\sum_{s>t} f^{ij}_{st} |\phi_{st}^{(-)} \rangle \langle \phi_{st}^{(-)} | \bigr)
&=
\frac{1}{d^2}\bigl( \mathbb{I}- \mathbb{F} \bigr) +
\frac{1}{d} \sum_{k>l}f^{ij}_{kl} |\phi_{kl}^{(-)} \rangle \langle \phi_{kl}^{(-)} |. \label{Eq:fspsi}
\end{align}
Hence, to obtain $\mc{R}^\ell(|\phi_{ij}^{(+)} \rangle \langle \phi_{ij}^{(+)} |)$ and $\mc{R}^\ell(|\phi_{ij}^{(-)} \rangle \langle \phi_{ij}^{(-)} |)$, we set
\begin{align}
&\mc{R}^\ell(|\phi_{ij}^{(+)} \rangle \langle \phi_{ij}^{(+)} |) = a^{(+)}_\ell (\mathbb{I} +\mathbb{F}) + b^{(+)}_\ell  \mathbb{L} + c^{(+)}_\ell \sum_{k>l} f_{kl}^{ij} |\phi_{kl}^{(+)} \rangle \langle \phi_{kl}^{(+)} |\\
&\mc{R}^\ell(|\phi_{ij}^{(-)} \rangle \langle \phi_{ij}^{(-)} |) = a^{(-)}_\ell (\mathbb{I} -\mathbb{F}) + c^{(-)}_\ell \sum_{k>l} f_{kl}^{ij} |\phi_{kl}^{(-)} \rangle \langle \phi_{kl}^{(-)} |,
\end{align}
and derive the coefficients using their recurrence relations.
From Eqs.~\eqref{Eq:R1symantisym0} and~\eqref{Eq:R1symantisym}, the coefficients for $n=1$ are given by
\begin{align}
&a_1^{(+)} = \frac{1}{d^2}\bigl( 1- \frac{2}{d} \bigr), \hspace{3mm}  b^{(+)}_1 = \frac{2}{d^3}, \hspace{3mm} c^{(+)}_1=\frac{1}{d},\\
&a_1^{(-)} = \frac{1}{d^2}, \hspace{3mm}  c^{(-)}_1=\frac{1}{d}.
\end{align}
From Eqs.~\eqref{Eq:R1symantisym0},~\eqref{Eq:R1symantisym},~\eqref{Eq:fsphi}, and~\eqref{Eq:fspsi},
recurrence relations for $a_\ell^{(\pm)}$, $q_\ell^{(+)}$, and $c_\ell^{(\pm)}$ are given by
\begin{align}
a_{\ell+1}^{(+)} = a_\ell^{(+)} + \frac{1}{d}\bigl(1 - \frac{1}{d} \bigr)q_\ell^{(+)} +\frac{1}{d^2}\bigl(1 - \frac{2}{d} \bigr)c_\ell^{(+)}, \hspace{3mm}
b_{\ell+1}^{(+)} =  \frac{q_\ell^{(+)} }{d^2} + \frac{2c_\ell^{(+)} }{d^3}, \hspace{3mm}
c_{\ell+1}^{(+)} =  \frac{c_\ell^{(+)} }{d},
\end{align}
and
\begin{align}
a_{\ell+1}^{(-)} = a_\ell^{(+)} + \frac{c_\ell^{(-)}}{d^2}, \hspace{3mm}
c_{\ell+1}^{(-)} =  \frac{c_\ell^{(-)} }{d}.
\end{align}
Solving these relations, we obtain
\begin{align}
a_{\ell}^{(+)} =\frac{1}{d(d+1)}- \frac{d^{\ell+1} + d^\ell - 2}{d^{2\ell+1} (d^2-1)}, \hspace{3mm}
q_{\ell}^{(+)} =  \frac{2(d^\ell-1) }{d^{2\ell+1}(d-1)} , \hspace{3mm}
c_{\ell}^{(+)} =d^{-\ell},
\end{align}
and
\begin{align}
a_{\ell}^{(-)} =\frac{1-d^{-\ell}}{d(d-1)}, \hspace{3mm}
c_{\ell}^{(-)} =d^{-\ell}.
\end{align}
Thus, we have
\begin{align}
&\mc{R}^\ell(|\phi_{ij}^{(+)} \rangle \langle \phi_{ij}^{(+)} |) = \biggl( 1- \frac{d^{\ell+1} + d^\ell - 2}{d^{2\ell} (d-1)} \biggr)\Pi^{\rm sym}+2 \frac{d^\ell-1}{d^{2\ell} (d-1)}\Lambda
+\frac{1}{d^\ell} \sum_{k>l}f^{i j}_{k l} |\phi_{kl}^{(+)} \rangle \langle \phi_{kl}^{(+)} |\\
&\mc{R}^\ell(|\phi_{ij}^{(-)} \rangle \langle \phi_{ij}^{(-)} |) = \bigl(1-\frac{1}{d^{\ell}} \bigr) \Pi^{\rm anti} 
+\frac{1}{d^\ell}\sum_{ k>l} f^{ij}_{kl} |\phi_{kl}^{(-)} \rangle \langle \phi_{kl}^{(-)} |.
\end{align}
This concludes the proof. $\hfill \blacksquare$
\end{Proof}

\bibliographystyle{unsrt}
\bibliography{Bib}

\begin{thebibliography}{10}

\bibitem{HaydenTutorial}
P.~Hayden.
\newblock Decoupling: A building block for quantum information theory.
\newblock \url{http://qip2011.quantumlah.org/images/QIPtutorial1.pdf}, 2012.
\newblock Accessed: 2017-3-30.

\bibitem{DBWR2010}
F.~Dupuis, M.~Berta, J.~Wullschleger, and R.~Renner.
\newblock One-shot decoupling.
\newblock {\em Commun. Math. Phys.}, 328:251, 2014.

\bibitem{SDTR2013}
O.~Szehr, F.~Dupuis, M.~Tomamichel, and R.~Renner.
\newblock Decoupling with unitary approximate two-designs.
\newblock {\em New J. Phys.}, 15:053022, 2013.

\bibitem{HM14}
C.~Hirche and C.~Morgan.
\newblock Efficient achievability for quantum protocols using decoupling
  theorems.
\newblock In {\em Proc. 2014 IEEE Int. Symp. Info. Theory}, page 536, 2014.

\bibitem{D2005}
I.~Devetak.
\newblock The private classical capacity and quantum capacity of a quantum
  channel.
\newblock {\em IEEE Trans. Inf. Theory}, 51(1):44--55, 2005.

\bibitem{DW2004}
I.~Devetak and A.~Winter.
\newblock Relating {Quantum} {Privacy} and {Quantum} {Coherence}: {An}
  {Operational} {Approach}.
\newblock {\em Phys. Rev. Lett.}, 93(8):080501, 2004.

\bibitem{HHWY2008}
P.~Hayden, M.~Horodecki, A.~Winter, and J.~Yard.
\newblock A decoupling approach to the quantum capacity.
\newblock {\em Open Syst. Inf. Dyn.}, 15:7, 2008.

\bibitem{ADHW2009}
A.~Abeyesinghe, I.~Devetak, P.~Hayden, and A.~Winter.
\newblock The mother of all protocols : Restructuring quantum information's
  family tree.
\newblock {\em Proc. R. Soc. A}, 465:2537, 2009.

\bibitem{DH2011}
N.~Datta and M.-H. Hsieh.
\newblock The apex of the family tree of protocols : optimal rates and resource
  inequalities.
\newblock {\em New J. Phys.}, 13:093042, 2011.

\bibitem{EAZ2005}
J.~Emerson, R.~Alicki, and K.~{\.Z}yczkowski.
\newblock Scalable noise estimation with random unitary operators.
\newblock {\em J. Opt. B: Quantum semiclass. opt.}, 7:S347--S352, 2005.

\bibitem{KLRetc2008}
E.~Knill, D.~Leibfried, R.~Reichle, J.~Britton, R.~B. Blakestad, J.~D. Jost,
  C.~Langer, R.~Ozeri, S.~Seidelin, and D.~J. Wineland.
\newblock Randomized benchmarking of quantum gates.
\newblock {\em Phys. Rev. A}, 77(1):012307, 2008.

\bibitem{MGE2011}
E.~Magesan, J.~M. Gambetta, and J.~Emerson.
\newblock Scalable and {Robust} {Randomized} {Benchmarking} of {Quantum}
  {Processes}.
\newblock {\em Phys. Rev. Lett.}, 106(18):180504, 2011.

\bibitem{MGE2012}
E.~Magesan, J.~M. Gambetta, and J.~Emerson.
\newblock Characterizing quantum gates via randomized benchmarking.
\newblock {\em Phys. Rev. A}, 85(4):042311, 2012.

\bibitem{PSW2006}
S.~Popescu, A.~J. Short, and A.~Winter.
\newblock Entanglement and the foundations of statistical mechanics.
\newblock {\em Nat. Phys.}, 2(11):754--758, 2006.

\bibitem{GLTZ2006}
S.~Goldstein, J.~L. Lebowitz, R.~Tumulka, and N.~Zangh\'{i}.
\newblock Canonical {Typicality}.
\newblock {\em Phys. Rev. Lett.}, 96(5):050403, 2006.

\bibitem{R2008}
P.~Reimann.
\newblock Foundation of {Statistical} {Mechanics} under {Experimentally}
  {Realistic} {Conditions}.
\newblock {\em Phys. Rev. Lett.}, 101(19):190403, 2008.

\bibitem{HP2007}
P.~Hayden and J.~Preskill.
\newblock Black holes as mirrors: quantum information in random subsystems.
\newblock {\em J. High Energy Phys.}, 2007(09):120, 2007.

\bibitem{SS2008}
Y.~Sekino and L.~Susskind.
\newblock Fast scramblers.
\newblock {\em J. High Energy Phys.}, 2008(10):065, 2008.

\bibitem{LSHOH2013}
N.~Lashkari, D.~Stanford, M.~Hastings, T.~Osborne, and P.~Hayden.
\newblock Towards the fast scrambling conjecture.
\newblock {\em J. High Energy Phys.}, 2013(4), 2013.

\bibitem{SS2014}
H.~Shenker and D.~Stanfor.
\newblock Black holes and the butterfly effect.
\newblock {\em J. High Energy Phys.}, 2014(3):67, 2014.

\bibitem{RD2015}
D.~A. Roberts and D.~Stanford.
\newblock Diagnosing {Chaos} {Using} {Four}-{Point} {Functions} in
  {Two}-{Dimensional} {Conformal} {Field} {Theory}.
\newblock {\em Phys. Rev. Lett.}, 115(13):131603, 2015.

\bibitem{SS2015}
S.~H. Shenker and D.~Stanford.
\newblock Stringy effects in scrambling.
\newblock {\em J. High Energy Phys.}, 2015(5):132, 2015.

\bibitem{DLT2002}
D.~P. DiVincenzo, D.~W. Leung, and B.~M. Terhal.
\newblock Quantum data hiding.
\newblock {\em IEEE Trans. Inf. Theory}, 48:580, 2002.

\bibitem{DCEL2009}
C.~Dankert, R.~Cleve, J.~Emerson, and E.~Livine.
\newblock Exact and approximate unitary 2-designs and their application to
  fidelity estimation.
\newblock {\em Phys. Rev. A}, 80:012304, 2009.

\bibitem{GAE2007}
D.~Gross, K.~Audenaert, and J.~Eisert.
\newblock Evenly distributed unitaries: {On} the structure of unitary designs.
\newblock {\em J. of Math. Phys.}, 48(5):052104, 2007.

\bibitem{TGJ2007}
G.~T\'{o}th and J.~J. Garc\'{i}a-Ripoll.
\newblock Efficient algorithm for multiqudit twirling for ensemble quantum
  computation.
\newblock {\em Phys. Rev. A}, 75(4):042311, 2007.

\bibitem{BWV2008a}
W.~G. Brown, Y.~S. Weinstein, and L.~Viola.
\newblock Quantum pseudorandomness from cluster-state quantum computation.
\newblock {\em Phys. Rev. A}, 77(4):040303(R), 2008.

\bibitem{WBV2008}
Y.~S. Weinstein, W.~G. Brown, and L.~Viola.
\newblock Parameters of pseudorandom quantum circuits.
\newblock {\em Phys. Rev. A}, 78(5):052332, 2008.

\bibitem{HL2009}
A.~W. Harrow and R.~A. Low.
\newblock Random quantum circuits are approximate 2-designs.
\newblock {\em Commun. Math. Phys.}, 291:257, 2009.

\bibitem{DJ2011}
I.~T. Diniz and D.~Jonathan.
\newblock {Comment on ``Random quantum circuits are approximate 2-designs"}.
\newblock {\em Commun. Math. Phys.}, 304:281, 2011.

\bibitem{HL2009TPE}
A.~W. Harrow and R.~A. Low.
\newblock Efficient {Quantum} {Tensor} {Product} {Expanders} and k-{Designs}.
\newblock In {\em Proc. RANDOM'09}.

\bibitem{BHH2012}
F.~G. S.~L. Brand\~{a}o, A.~W. Harrow, and M.~Horodecki.
\newblock Local random quantum circuits are approximate polynomial-designs.
\newblock arXiv:1208.0692, 2012.

\bibitem{CLLW2015}
R.~Cleve, D.~Leung, L.~Liu, and C.~Wang.
\newblock Near-linear constructions of exact unitary 2-designs.
\newblock {\em Quant. Info. {\&} Comp.}, 16(9 {\&} 10):0721--0756, 2016.

\bibitem{L2010}
R.~A. Low.
\newblock {\em Pseudo-randomness and learning in quantum computation}.
\newblock PhD thesis, University of Bristol, 2010.
\newblock arXiv:1006.5227.

\bibitem{AG2004}
S.~Aaronson and D.~Gottesman.
\newblock Improved simulation of stabilizer circuits.
\newblock {\em Phys. Rev. A}, 70(5):052328, 2004.

\bibitem{RLL2009}
C.~A. Ryan, M.~Laforest, and R.~Laflamme.
\newblock Randomized benchmarking of single- and multi-qubit control in
  liquid-state {NMR} quantum information processing.
\newblock {\em New J. Phys.}, 11(1):013034, 2009.

\bibitem{BWCetc2011}
K.~R. Brown, A.~C. Wilson, Y.~Colombe, C.~Ospelkaus, A.~M. Meier, E.~Knill,
  D.~Leibfried, and D.~J. Wineland.
\newblock Single-qubit-gate error below {$10^{-4}$} in a trapped ion.
\newblock {\em Phys. Rev. A}, 84(3):030303, 2011.

\bibitem{CGJetc2013}
A.~D. C\'{o}rcoles, Jay~M. Gambetta, Jerry~M. Chow, John~A. Smolin, Matthew
  Ware, Joel Strand, B.~L.~T. Plourde, and M.~Steffen.
\newblock Process verification of two-qubit quantum gates by randomized
  benchmarking.
\newblock {\em Phys. Rev. A}, 87(3):030301, 2013.

\bibitem{BKMetc2014}
R.~Barends, J.~Kelly, A.~Megrant, A.~Veitia, D.~Sank, E.~Jeffrey, T.~C. White,
  J.~Mutus, A.~G. Fowler, B.~Campbell, Y.~Chen, Z.~Chen, B.~Chiaro,
  A.~Dunsworth, C.~Neill, P.~O’Malley, P.~Roushan, A.~Vainsencher, J.~Wenner,
  A.~N. Korotkov, A.~N. Cleland, and John~M. Martinis.
\newblock Superconducting quantum circuits at the surface code threshold for
  fault tolerance.
\newblock {\em Nature}, 508(7497):500--503, 2014.

\bibitem{NHMW2015-2}
Y.~Nakata, C.~Hirche, C.~Morgan, and A.~Winter.
\newblock Decoupling with random diagonal unitaries.
\newblock arXiv:1509.05155, 2015.

\bibitem{KSV2002}
A.~Kitaev, A.~Shen, and M.~Vyalyi.
\newblock {\em {\it Classical and Quantum Computation}}.
\newblock American Mathematical Society Boston, MA, USA, 2002.

\bibitem{M1990}
M.~L. Metha.
\newblock {\em {\it Random Matrices}}.
\newblock Academic Press, Amsterdam San Diego Oxford London, 1990.

\bibitem{NM2013}
Y.~Nakata and M.~Murao.
\newblock Diagonal-unitary 2-designs and their implementations by quantum
  circuits.
\newblock {\em Int. J. Quant. Inf.}, 11:1350062, 2013.

\bibitem{NTM2012}
Y.~Nakata, P.~S. Turner, and M.~Murao.
\newblock Phase-random states: {Ensembles} of states with fixed amplitudes and
  uniformly distributed phases in a fixed basis.
\newblock {\em Phys. Rev. A}, 86(1):012301, 2012.

\bibitem{NKM2014}
Y.~Nakata, M.~Koashi, and M.~Murao.
\newblock Generating a state t-design by diagonal quantum circuits.
\newblock {\em New J. Phys.}, 16:053043, 2014.

\bibitem{GR1999}
R.~Goodman and N.~R. Wallach.
\newblock {\em {\it Representations and Invariants of the Classical Groups}}.
\newblock Cambridge University Press, Cambridge, UK, 1999.

\bibitem{J1972}
A.~Jamio\l kowski.
\newblock Linear transformations which preserve trace and positive
  semidefiniteness of operators.
\newblock {\em Rep. Math. Phys.}, 3:275, 1972.

\bibitem{C1975}
M.~D. Choi.
\newblock Completely positive linear maps on complex matrices.
\newblock {\em Linear Algebra Appl.}, 10:285, 1975.

\bibitem{W2012}
M.~M. Wolf.
\newblock Quantum channels and operations: Guided tour.
\newblock
  \url{http://www-m5.ma.tum.de/foswiki/pub/M5/Allgemeines/MichaelWolf/QChannelLecture.pdf},
  2012.

\end{thebibliography}

\end{document}